\documentclass[showpacs,preprintnumbers,amsmath,amssymb,prl,aps,twocolumn,superscriptaddress]{revtex4}

\usepackage{graphicx}
\usepackage{bm}
\usepackage[usenames,dvipsnames]{color} \usepackage{ulem}

\newcommand{\bk}{{\bf k}} 
\newcommand{\bq}{{\bf q}}

\begin{document}
 
\date{\today} 
\title{The effects of non-linear electron-phonon interactions on 
superconductivity and charge-density-wave correlations}

\author{Shaozhi Li and S. Johnston}
\affiliation{Department of Physics and Astronomy, The University of Tennessee, Knoxville, TN 37996}
\begin{abstract}  
Determinant quantum Monte Carlo (DQMC) simulations are used to study non-linear
electron-phonon interactions in a two-dimensional Holstein-like model on
a square lattice. We examine the impact of non-linear electron-lattice
interactions on superconductivity and on Peierls charge-density-wave (CDW)
correlations at finite temperatures and carrier concentrations. We find that
the CDW correlations are dramatically suppressed with the inclusion of even a
small non-linear interaction. Conversely, the effect of the non-linearity  
on superconductivity is found to be less dramatic at high
temperatures; however, we find evidence that the non-linearity is ultimately
detrimental to superconductivity. These effects are attributed to the
combined hardening of the phonon frequency and a renormalization of the
effective linear electron-phonon coupling towards weaker values. These results
demonstrate the importance of non-linear interactions at finite carrier
concentrations when one is addressing CDW and superconducting order and have 
implications for experiments that drive the lattice far from equilibrium.
\end{abstract}
 
\pacs{02.70.Ss, 74.25.Kc, 71.38.-k} \maketitle 

Electron-phonon ($e$-ph) coupling is an important interaction in many molecular systems
and solids, which dresses carriers to form quasiparticles (called polarons) with
increased effective masses and modified dispersion relations \cite{Polaron,Kink}. 
Nearly all
treatments of this interaction make use of linear models, where an electronic
degree of freedom is coupled to the first order displacement of the ions.
Theorists usually justify this by expanding the interaction in powers of the
atomic displacement and then truncating the expansion to first order by
assuming that the net displacements are small. But large displacements are
expected in a number of situations. 
For example, in the limit of strong $e$-ph 
coupling, linear models predict large lattice distortions surrounding carriers
as small polarons are formed \cite{Polaron,AdolphsEPL2013,MA,AdolphsPRB2014}.  
This result violates the assumptions underlying
the linear models and indicates the necessity of including higher order terms
in the expansion \cite{AdolphsEPL2013,AdolphsPRB2014}.

Non-linear $e$-ph interactions have a dramatic effect on polaron properties in
the single carrier limit. This was first demonstrated in Ref. \onlinecite{AdolphsEPL2013}, which examined
the problem using the non-perturbative ``momentum average" approximation and
found that small higher-order interactions lead to a dramatic undressing of the
polaron. This result is of potential relevance to many systems where strong
$e$-ph interactions have been inferred from experiments \cite{Mat1,Mat2,Mat3,Mat4,Mat5,MgB2}; 
however, it is not clear whether the single polaron result will generalize
straightforwardly to the finite carrier concentrations relevant for these
materials. Calculations at finite carrier concentrations are also needed in order to understand
the impact of non-linearity on broken symmetry states like superconductivity \cite{SC}
and charge-density-waves (CDW) \cite{CDW}. 

Non-linearity can also be important when the $e$-ph interaction is weak. In some
systems small initial atomic displacements can be driven far from
equilibrium by an external potential. For instance, coherent phonon excitations
have been observed in several pump-probe experiments 
\cite{Time1,Time2,Time3,Time4,Time5}. These experiments offer a
promising new path to probing $e$-ph interactions, particularly in systems where
the phonons have been difficult to differentiate from other collective
excitations \cite{Time6}. If the pump fluence is too high, however, it is possible that the
lattice will be driven far from equilibrium, making higher-order terms in the $e$-ph
interaction relevant.

In this letter we examine non-linear interactions in the many-body limit by
studying the non-linear single-band Holstein model in two dimensions using
determinant quantum Monte Carlo (DQMC). DQMC is a non-perturbative
auxiliary-field technique capable of handling the $e$-ph interactions in a
numerically exact fashion \cite{WhitePRB1989,BSD}. The technique is formulated in the grand canonical
ensemble, which allows us to study the model at finite carrier concentrations
and temperatures. DQMC has previously been applied to linear Holstein
models \cite{ScalettarPRB1989,CreffieldEPJ2005,JohnstonPRB2013}, 
but to the best of our knowledge, it has not been applied to any $e$-ph
models with non-linear interactions. Here we focus on the competition between
Peierls CDW correlations and $s$-wave superconductivity that is known to occur in
the linear model \cite{ScalettarPRB1989,MarsiglioPRB1990}. 
As with the single carrier limit, we find that small
non-linear interactions alter the properties of the system at finite carrier
concentrations, undressing the carriers and significantly suppressing CDW
correlations. While this allows superconductivity to emerge from behind
the competing CDW order, we conclude that a non-linear interaction is
ultimately detrimental to superconductivity in the Holstein model due 
to a renormalization of the effective linear $e$-ph coupling.

We consider a modified single-band Holstein 
Hamiltonian $H = H_\mathrm{el} + H_\mathrm{lat} + H_\mathrm{int}$, where
\begin{equation}\nonumber
H_\mathrm{el}=-t\sum_{\langle i,j\rangle,\sigma} c^\dagger_{i,\sigma} c^{\phantom\dagger}_{j,\sigma} - 
\mu\sum_{i,\sigma} \hat{n}_{i,\sigma}, 
\end{equation}
\begin{equation}\nonumber
H_\mathrm{lat}=\sum_i \left(
\frac{1}{2M}\hat{P}^2_i +\frac{M\Omega^2}{2}\hat{X}_i^2  
\right) = 
\sum_i \Omega \left(b^\dagger_ib^{\phantom\dagger}_i+\frac{1}{2}\right) 
\end{equation}
\begin{equation}
H_\mathrm{int}=\sum_{i,\sigma,k} \alpha_k \hat{n}_{i,\sigma} \hat{X}^k_i 
= \sum_{i,\sigma,k}g_k\hat{n}_{i,\sigma}\left(b^\dagger_i + b^{\phantom\dagger}_i\right)^k.
\end{equation}
Here, $c^\dagger_{i,\sigma}$ ($c^{\phantom\dagger}_{i,\sigma}$) creates
(annihilates) an electron of spin $\sigma$ on lattice site $i$; 
$b^\dagger_i$ ($b_i$) creates (annihilates) a phonon on site $i$;  
$\hat{n}^{\phantom\dagger}_{i,\sigma} = c^\dagger_{i,\sigma}c^{\phantom\dagger}_{i,\sigma}$
is the electron number operator; $t$ is the nearest-neighbor hopping integral;
$\mu$ is the chemical potential which sets the band filling;
$\hat{X}_i$ and $\hat{P}_i$ are the position and momentum operators, respectively,
for lattice site $i$; $\Omega$ is the phonon frequency; $\langle \dots\rangle$
denotes a sum over nearest neighbors; and $g_k = \alpha_k(2M\Omega)^{-\frac{k}{2}}$ 
is the strength of the $e$-ph coupling to $k$-th order in displacement.
Throughout this work consider a two-dimensional square lattice with lattice spacing $a$ 
and set $a = M = t = 1$ as the units of length,
mass, and energy, respectively.

It is convenient to define a 
dimensionless $e$-ph coupling strength $\lambda$, which is given by the ratio of the
lattice deformation energy $E_p$ to half the electronic bandwidth $W/2$.
For the linear Holstein Hamiltonian in two-dimensions 
$\lambda = \alpha_1^2/(M\Omega^2 W) = g_1^2/4t\Omega$. 
For the non-linear model, additional dimensionless ratios 
$\xi_n = g_n/g_{n-1}$ must also be specified in order
characterize the strength of the non-linear terms.  
In the single carrier limit the quartic terms have
a much weaker effect on the properties of the polaron in comparison to the
quadratic terms \cite{AdolphsEPL2013}. We expect a similar result here. We therefore 
limit ourselves to the linear and quadratic order couplings only and define $\xi = g_2/g_1$. 
We note that one might be tempted to return to the 
physical definition of $\lambda$ in order to characterize the system with
a single effective parameter. The implicit assumption here is that
the system can be mapped onto an effective linear model with a renormalized
phonon frequency $\Omega$ and $e$-ph coupling $g$.
This, however, is not possible for the single particle
case \cite{AdolphsEPL2013}, where such effective linear models fail to capture
the results of the non-linear model. Therefore multiple parameters are needed to
characterize the non-linear model. In order to facilitate
easy comparisons with the linear case we keep the standard definition of 
$\lambda$, where $\lambda > 1$ implies strong linear coupling, and
use the ratio $\xi$ to characterize the strength of the non-linearity. Other choices
are possible.

We study the non-linear model using DQMC. The method is outlined in a number of 
references \cite{BSD,WhitePRB1989}, and complete details of its application to the lattice
degrees of freedom can be found in Ref. \onlinecite{JohnstonPRB2013}. The only
change from the procedure outlined therein is with regards to the definition of
the $B$-matrices, which are defined on each discrete time slice $l$ [Eq. (11) of
Ref. \onlinecite{JohnstonPRB2013}]; they must be modified to include the 
higher order interaction terms. Specifically,  
$B_\sigma(l) = \exp(-\Delta\tau v(l))\exp(-\Delta\tau K)$, 
where $K$ is the matrix representation of $H_\mathrm{el}$ and $v(l)$ is a diagonal matrix
whose $i$-th element is $[v(l)]_{ii} = (\sum_k \alpha_kX^{k}_{i,l})$. All other aspects of
the problem, including the sampling of the phonon fields, are treated as
described in Ref. \onlinecite{JohnstonPRB2013}. 
Throughout this work we performed simulations 
with $\Delta\tau = 1/10$ and generally work with clusters $N = 8\times 8$ in size, 
but some results are shown for other cluster sizes.  
We have examined clusters with linear dimensions ranging from $N = 4$ to $12$ and 
various values of $\Delta\tau$ for the half-filled case and found 
no significant finite size effects or $\Delta\tau$ errors. Finally, we note that both the 
linear and non-linear Holstein models do not exhibit a Fermion sign problem, which 
allows us to perform simulations to low temperatures.

We first examine the CDW and superconducting correlations. A 
measure of the CDW correlations is obtained from the charge susceptibility  
\begin{equation}
\chi_C(\bq) = \frac{1}{N}\int_0^\beta d\tau 
\langle \rho(\bq,\tau)\rho^\dagger(\bq,0)\rangle, 
\end{equation}
where $\rho({\bf q}) = \sum_{i,\sigma} e^{i\bq\cdot {\bf R}_i} \hat{n}_{i,\sigma}$. 
Similarly, a measure of the $s$-wave superconducting correlations is obtained
from the pair-field susceptibility
\begin{equation}
\chi_{SC} = \frac{1}{N}\int_0^\beta d\tau 
\langle \Delta(\tau)\Delta^\dagger(0)\rangle,  
\end{equation}
where $\Delta^\dagger = \sum_{\bk} c^\dagger_{\bk,\uparrow}c^\dagger_{-\bk,\downarrow} 
= \sum_i c^\dagger_{i,\uparrow} c^\dagger_{i,\downarrow}$.

\begin{figure}
\includegraphics[width=\columnwidth]{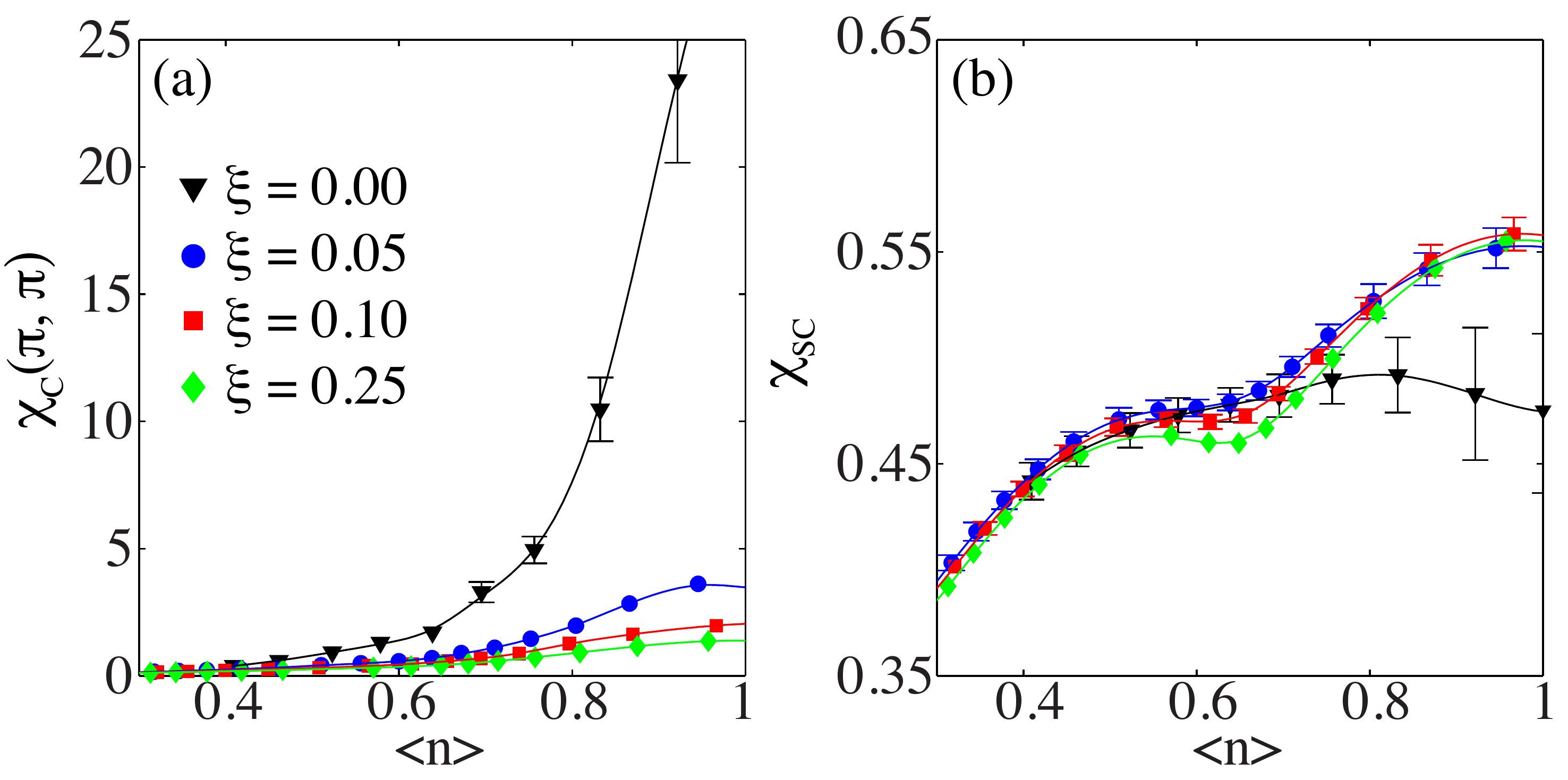}
\caption{(color online) 
 (a) $\bq = (\pi,\pi)$ charge and (b) pair-field susceptibilities
 as a function of filling for the linear ($\xi = 0$, black downward triangle)
 and non-linear ($\xi = 0.05$, blue circle;
 $\xi = 0.10$, red square; and $\xi = 0.25$,
 green diamond) Holstein models with $\Omega = t$ and $\lambda = 0.25$.
 The remaining parameters are $N = 8\times 8$, $\beta = 5/t$,
 and $\Delta\tau = 0.1/t$.
 Error bars smaller than the marker size
 have been suppressed for clarity.
}
\label{Fig:1}
\end{figure}

Fig. \ref{Fig:1} shows the CDW and superconducting correlations 
as a function of the band filling in the linear and non-linear models at an inverse
temperature $\beta = 5/t$. Here the linear coupling has been fixed to
$\lambda = 0.25$. The results for the linear model ($\xi = 0$) agree well with previous
work \cite{ScalettarPRB1989}, where $\bq = (\pi,\pi)$ CDW correlations dominant 
due to a strong $\bq = (\pi,\pi)$ nesting condition on the Fermi surface 
near half-filling ($\langle n\rangle\sim 1$).\cite{ScalettarPRB1989,MarsiglioPRB1990} 
The non-linear interaction dramatically alters these results. The initial
effect is rapid and we find that
$\chi_C(\pi,\pi)$ is suppressed near $\langle n \rangle \sim 1$
by an order of magnitude for a relatively small
value of the non-linear coupling $\xi = 0.05$.
This suppression continues for increasing values of $\xi$, but it is 
less dramatic after the initial decrease.

\begin{figure}[t!]
 \begin{center}
 \includegraphics[width=\columnwidth]{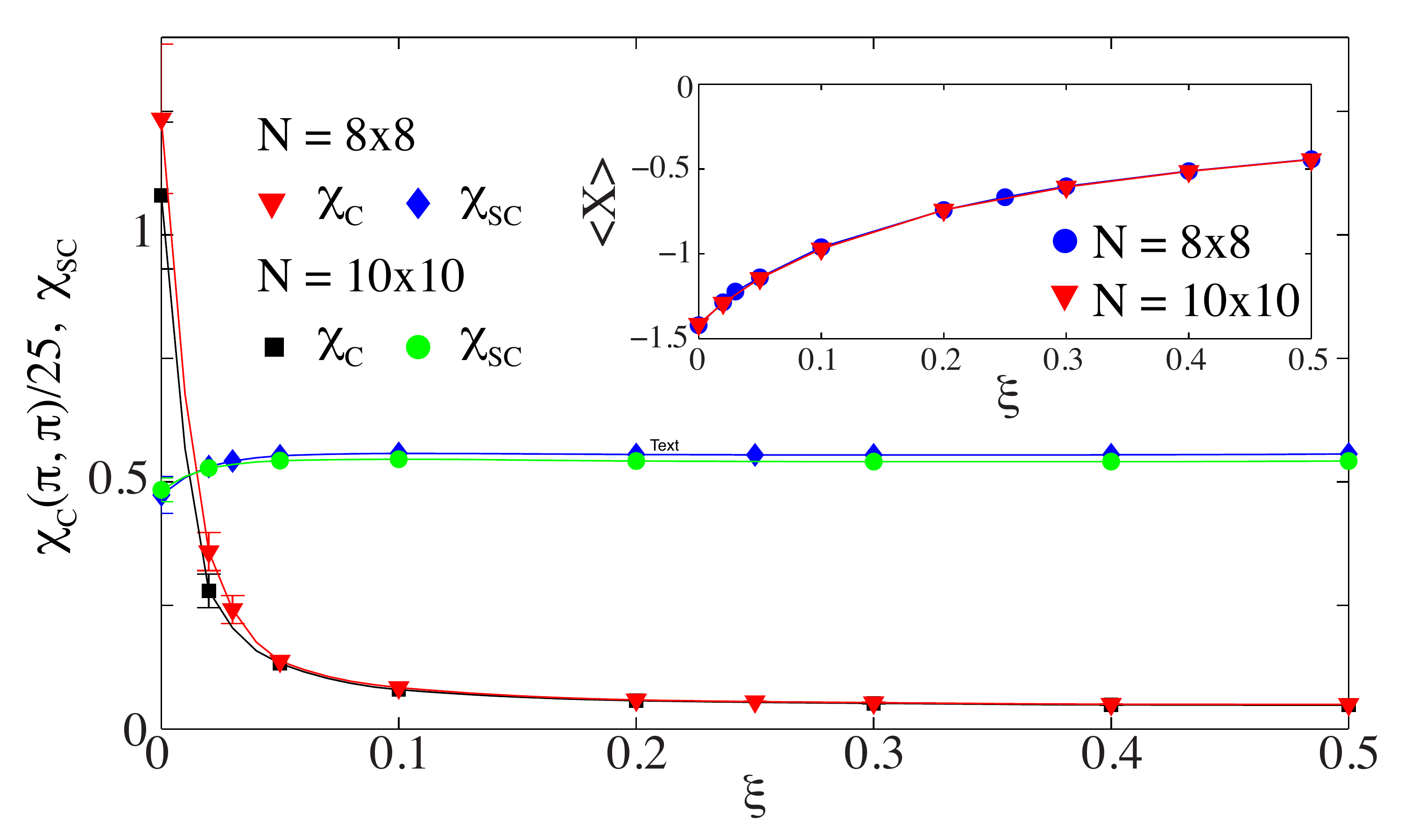}
 \end{center}
 \caption{\label{Fig:2} (color online) The charge $\chi_C(\pi,\pi)$
 (red downward triangle and blue diamond)
 and pair-field $\chi_{SC}$ (black square and green circle)
 susceptibilities as a function of $\xi$ at
 half filling $\langle n \rangle = 1$. Results are shown for $N = 8\times8$ and $10\times10$
 clusters while the remaining parameters
 are the same as those used in Fig. \ref{Fig:1}. The charge susceptibility
 has been rescaled by a factor of $25$. Inset: The average value of the
 lattice displacement as a function of $\xi$ for both cluster sizes.
 Error bars smaller than the marker size have been suppressed for clarity.}
\end{figure}

The $\xi$ dependence of $\chi_{C}(\pi,\pi)$ and $\chi_{SC}$ is examined
further in Fig. \ref{Fig:2}. Results are shown for a fixed filling of
$\langle n \rangle = 1$ and $\lambda = 0.25$. The behavior matches the expectations from
Fig. \ref{Fig:1} and the rapid initial suppression of the CDW
correlations for small non-zero values of $\xi$ is evident.
Similar results were obtained in the single polaron limit, where a
small value of $\xi$ produced large changes in the polaron's effective mass and
quasiparticle weight, but gave way to more gradual changes
in these properties for further increases in the value of $\xi$ \cite{AdolphsEPL2013}.
We also note that the results obtained for $N = 8\times 8$ and $N = 10\times 10$ clusters 
are nearly identical, indicating that the finite size effects are small and 
have little bearing on our conclusions. 

In the linear model, CDW correlations directly compete with
$s$-wave superconductivity and the former dominate at low temperatures,
particularly for fillings near $\langle n \rangle \sim 1$
\cite{ScalettarPRB1989,MarsiglioPRB1990}.
Thus there is a concomitant enhancement in the pair-field susceptibility
once the CDW correlations are suppressed by the non-linear interaction, which is
evident in Figs. \ref{Fig:1} and \ref{Fig:2}. After its initial rise, however, $\chi_{SC}$ is
relatively independent of the value of $\xi$ for all values of the band
filling examined, apart from a slight suppression of $\chi_{SC}$ 
in the vicinity of $\langle n\rangle \sim 0.65$. Thus the non-linear coupling
does not significantly enhance or suppress superconductivity at this temperature 
once the competition with the CDW correlations has been suppressed or eliminated.

The inset of Fig. \ref{Fig:2} plots the average lattice displacement
$\langle X \rangle = \frac{1}{N}\sum_i X_i$ as a function of $\xi$. This quantity serves as a
proxy for the average number of phonon quanta (which is not directly accessible in the DQMC formalism)
as larger lattice distortions are described by coherent states with increasing numbers of
phonon quanta. For increasing values of $\xi$, the average lattice displacement is reduced,
and thus, so is the number of phonon quanta on each site. This is fully consistent
with the single carrier limit where the number of phonon quanta in the polaron cloud
dropped dramatically for non-zero values of $\xi$ \cite{AdolphsEPL2013}.
This relaxation of the lattice displacement shown here thus 
reflects the undressing of the lattice bipolarons that form the ${\bf q} = (\pi,\pi)$ CDW state.  

The physical origin of the polaron undressing is the 
renormalization of the effective linear coupling 
by the non-linear interaction terms.  
This can be understood from a mean-field treatment of the problem in the 
atomic limit \cite{AdolphsEPL2013}. Applying the mean-field decoupling 
$(b^\dagger)^2 = 2b^\dagger \langle b^\dagger\rangle - \langle b^\dagger\rangle^2$ 
leads to the effective Hamiltonian 
\begin{equation} 
H_{MF} = \Omega_{MF} \left(b^\dagger b +\frac{1}{2}\right)+ g_{MF}\hat{n}(b^\dagger + b)
\end{equation}
where $\Omega_{MF} = \Omega + 2g_2$ and $g_{MF} = g_1\left(1-\frac{2g_2}{\Omega+4g_2}\right)$. 
From this result one sees that the second order interaction acts to  
harden the phonon frequency and renormalize the effective linear coupling to lower 
values. This has the net effect of {\it decreasing} $\lambda_{MF}\sim \frac{g_{MF}^2}{\Omega_{MF}}$ and 
thus the effective $e$-ph interaction is weaker for the non-linear model. A similar effect 
occurs for the itinerant case examined here, as is evident from the relaxation 
of the average lattice displacement and the suppression of the CDW correlations.   
This observation also helps us understand why superconductivity is not strongly 
affected. While the hardening of the phonon frequency would raise the superconducting 
$T_c$, the net decrease in effective linear coupling results in an overall suppression of 
of the superconducting correlations and thus $T_c$. 

If the non-linear coupling results in a undressing of the polarons via a 
weakening of the effective linear coupling, we would expect the system to relax 
back to a metallic state for large values of $\xi$.  
We therefore examine the spectral weight at the Fermi level order to confirm 
this expectation. 
A measure of the spectral weight at the Fermi level 
($\omega = 0$) can be obtained from the imaginary time Green's function via the
relationship
\cite{TrivediPRL1995}
\begin{equation}
\beta G(\bk,\tau=\beta/2) = \frac{\beta}{2}\int d\omega 
A(\bk,\omega)g(\omega,\beta),  
\end{equation}
where $g(\omega,\beta) = \cosh^{-1}(\beta\omega/2)$
and $A(\bk,\omega) = -\frac{1}{\pi}{\mathrm{Im}}G(\bk,\omega)$ is the spectral function.
At low temperatures $g(\omega,\beta)$ is peaked at $\omega = 0$ and
therefore weights the spectral weight at the Fermi level.
The local propagator $G({\bf r}-{\bf r}^\prime,\tau=\beta/2) \propto \sum_{\bk} G(\bk,\beta/2)$
is then a measure of the total spectral weight at the Fermi level.
For simplicity we introduce the notation $\beta G({\bf r} = {\bf r}^\prime,\tau=\beta/2) 
\equiv \beta G_{\beta/2}$.   

\begin{figure}
 \includegraphics[width=\columnwidth]{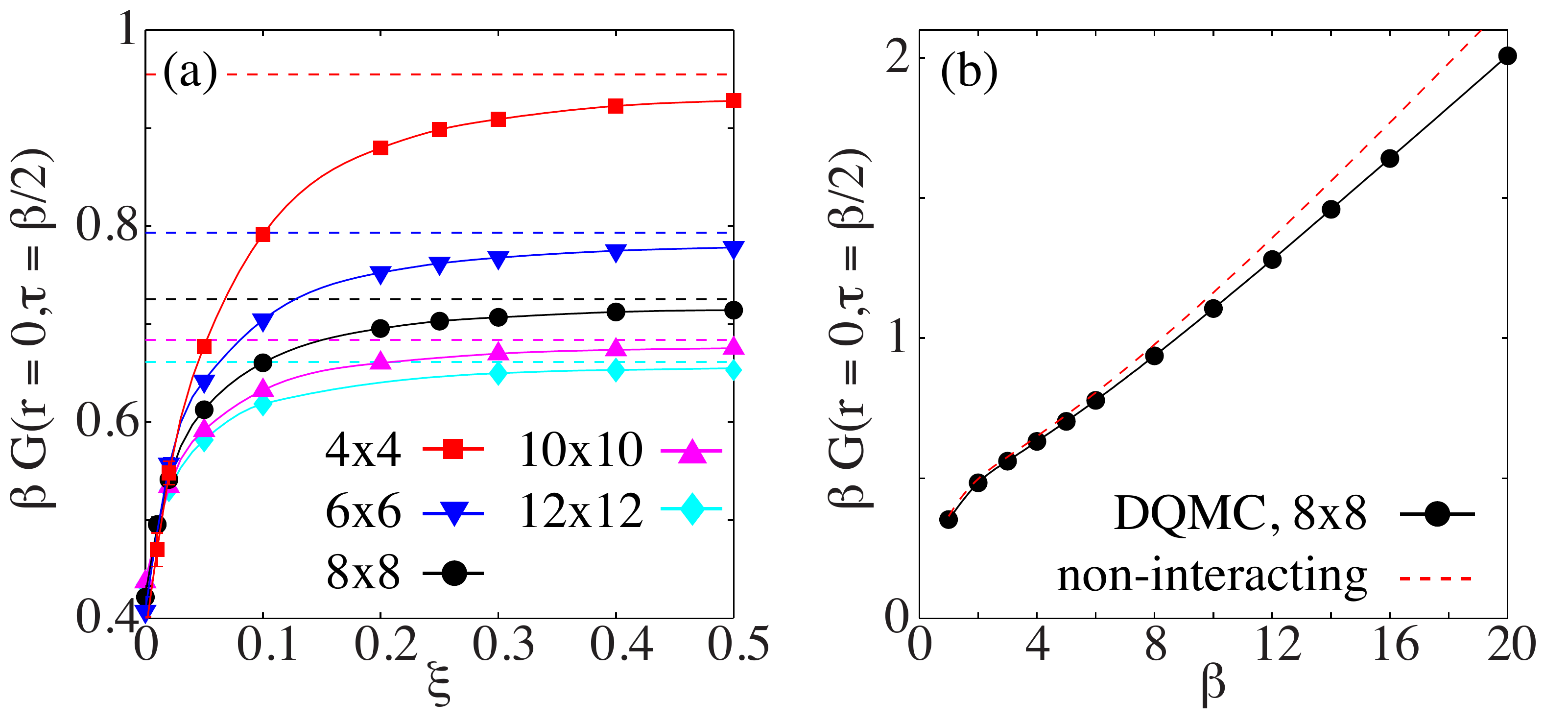}
 \caption{\label{Fig:3} (color online) (a) The spectral weight at the Fermi level given by
 $\beta G({\bf r} = 0,\tau=\beta/2) \equiv \beta G_{\beta/2}$ (see main text) as a function
 of the non-linear coupling $\xi$ for various cluster sizes, 
 $\Omega = t$, and $\lambda = 0.25$. The inverse temperature is $\beta = 5/t$.
 The dashed lines indicate the value for the non-interacting ($\lambda = \xi = 0$)
 case $\beta G^{(0)}_{\beta/2}$.
 (b) The temperature dependence of $\beta G_{\beta/2}$ for the $N = 8\times 8$ cluster 
 and $\xi = 0.25$. 
 The red dashed line is the result for the non-interacting case.
 Error bars smaller than the marker size
 have been suppressed for clarity.}
\end{figure}

Fig. \ref{Fig:3}a shows $\beta G_{\beta/2}$ as a function 
of $\xi$. Results are shown at half-filling for a number of cluster sizes 
and the behavior is similar for all cases examined. For $\xi = 0$ the CDW correlations dominate, 
resulting in the formation of a CDW gap that reduces the spectral weight at $\omega = 0$. 
The spectral weight is restored for increasing values of $\xi$, 
which is consistent with the closing of the CDW gap.   
For large $\xi$ the value of $\beta G_{\beta/2}$ approaches the 
non-interacting value, which is indicated by the dashed lines. 
Therefore, at $\beta = 5/t$, the system is metallic 
but with a slightly reduced $\xi$-dependent spectral weight. 
The metallicity of the system is further evidenced by the temperature dependence 
of $\beta G_{\beta/2}$, shown in Fig. \ref{Fig:3}b for the $\xi = 0.25$, 
$N = 8\times 8$ case. Here    
$\beta G_{\beta/2}$ increases for decreasing temperatures as expected 
for a Fermi liquid where the quasiparticle scattering rate scales as $T^2$.   
The full DQMC result, however, deviates from the non-interacting limit, 
indicating that the quasiparticles remain slightly dressed by the 
$e$-ph interaction. This picture is consistent with the one obtained 
from the single carrier limit, where the small polaron relaxes to a large polaron 
with a renormalization factor $Z$ only slightly reduced from 1  
at large values of $\xi$ \cite{AdolphsEPL2013}. 

\begin{figure}
 \begin{center}
 \includegraphics[width=\columnwidth]{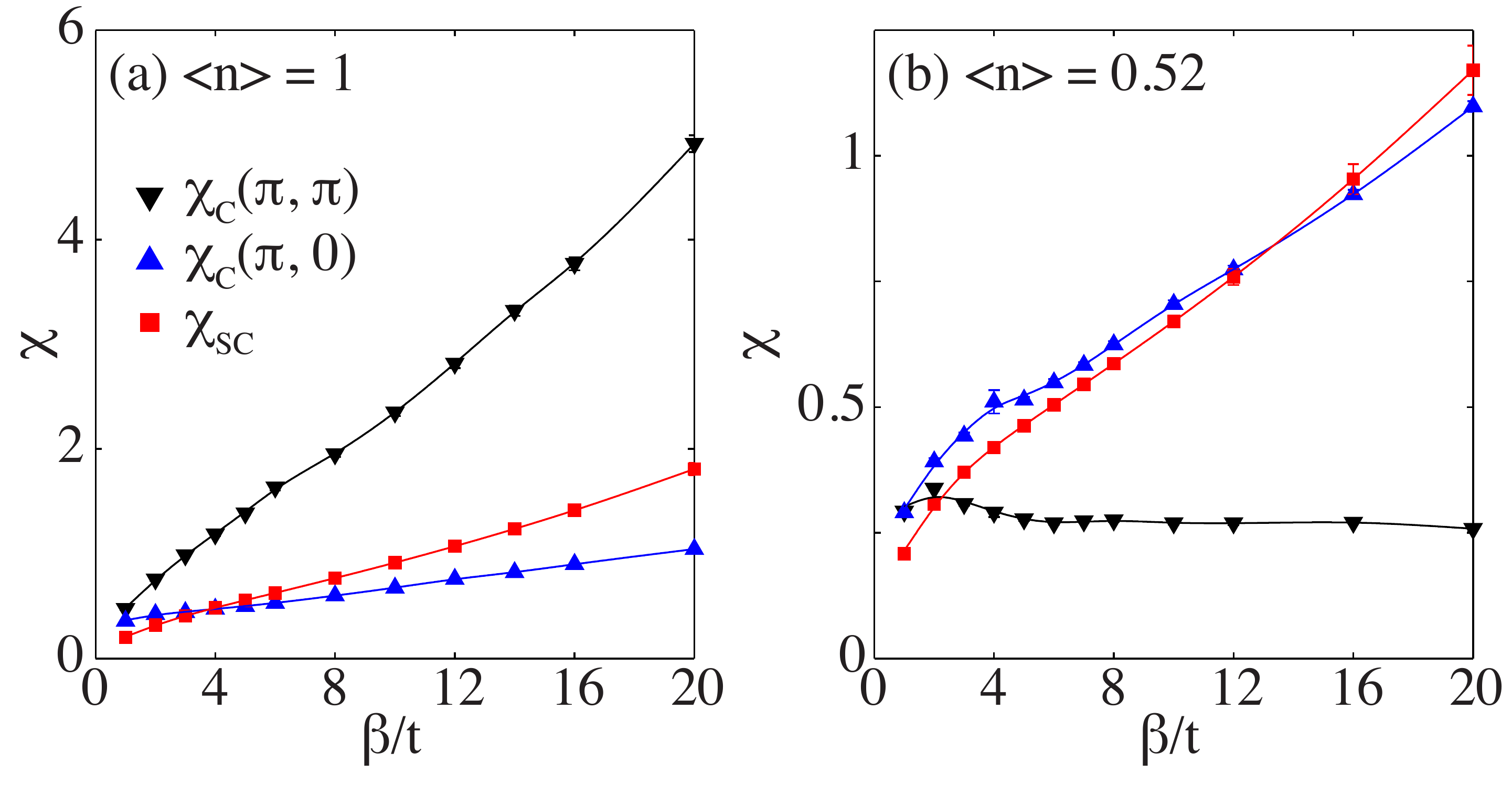}
 \end{center}
 \caption{\label{Fig:4} (color online) The charge $\chi_C(\bq)$ and 
 pair-field $\chi_{SC}$ susceptibilities as a function of the 
 inverse temperature for the 
 non-linear model ($\xi = 0.25$) at fillings (a) $\langle n \rangle = 1$ and (b) 
 $\langle n \rangle = 0.52$. The charge susceptibilities at 
 wavevectors $\bq = (\pi,\pi)$ (black downward triangle)  
 and $\bq = (\pi,0)$ (blue upward triangle) are shown. The remaining parameters 
 were $\lambda = 0.25$, $\Omega = t$, and $\Delta\tau = 0.1/t$.
 Error bars smaller than the marker size 
 have been suppressed for clarity.}
\end{figure}
 
We have demonstrated that the system re-enters a metallic phase 
as the value of $\xi$ is increased and the effective linear coupling is 
decreased. We would therefore like to assess if 
superconductivity emerges as the ground state if the temperature is lowered further. 
To examine this, Fig. \ref{Fig:4}a plots the temperature dependence of several 
relevant susceptibilities for the half- (Fig. \ref{Fig:4}a) 
and approximately quarter-filled models (Fig. \ref{Fig:4}b).   
At half-filling the $\bq = (\pi,\pi)$ CDW correlations are weakened, however, they 
remain as the dominate correlations in the system  
for all values of $\xi$ examined [for reference,  
$\chi_C(\pi,\pi)/\chi_{SC} \sim 2.2$ for $\xi = 0.5$ and $\beta = 5/t$]. 
This remains true upon further cooling and thus the 
ground state of the system with $\xi = 0.25$ remains 
a $\bq = (\pi,\pi)$ CDW insulator albeit with a 
drastically reduced transition temperature. 

Away from half-filling $\chi_{C}(\pi,\pi)$ is reduced 
by a combination of the non-linear interaction and 
the loss of the Fermi surface nesting at this wavevector. 
For example, in the vicinity of a 
quarter filling we find $\chi_{SC} > \chi_C(\pi,\pi)$. 
But other ordering vectors become relevant at these fillings, and for  
$\langle n \rangle\sim0.52$ we find that $\bq = (\pi,0)$ 
becomes the dominant ordering vector. Moreover, for $\beta = 5/t$,
$\chi_C(\pi,0) \sim \chi_{SC}$ suggesting that superconductivity could emerge
as the ground state at this filling. For decreasing temperatures  
$\chi_C(\pi,0)$ and $\chi_{SC}$ increase concurrently, but the 
superconducting pair-field susceptibility  
overtakes the charge susceptibility at $\beta \sim 14/t$. 
This signals a superconducting 
ground state at low temperature but with a reduced T$_c$ 
owing to the renormalized effective linear coupling.  

\begin{figure}
 \includegraphics[width=\columnwidth]{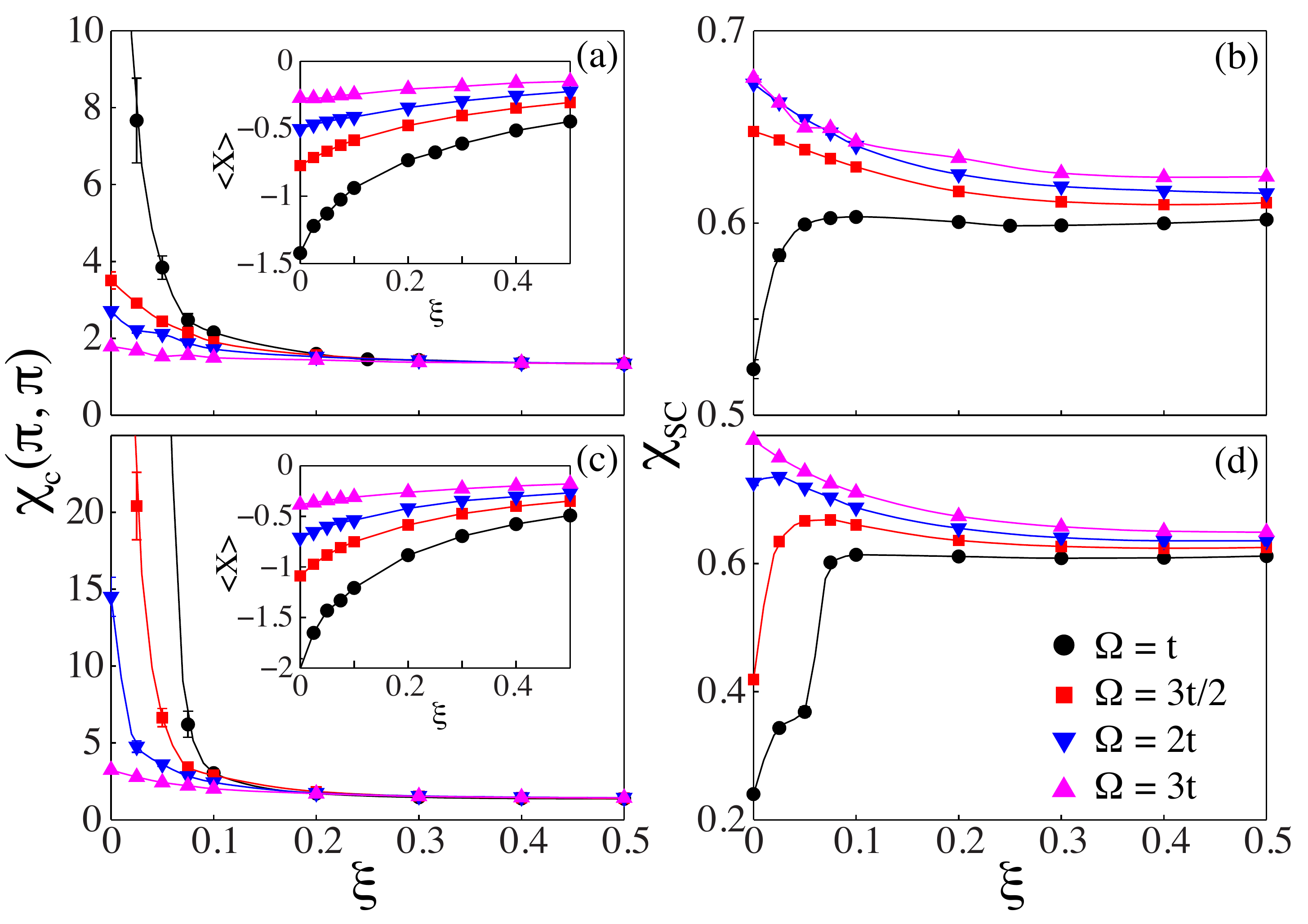}
 \caption{\label{Fig:5} (color online) The $\Omega$-dependence of the 
 (a), (c) charge $\chi_C(\pi,\pi)$ and (b), (d) superconducting pair 
 field $\chi_{SC}$ susceptibilities for the half-filled model. 
 The linear couplings are $\lambda = 0.25$ for the top panels and 
 $\lambda = 0.5$ for the bottom panels. Results are shown for an $N = 8\times 8$ 
 cluster. Error bars smaller than the marker size
 have been suppressed for clarity.
 }
\end{figure}

Finally, in Fig \ref{Fig:5} we examine the dependence of our results at 
half-filling on the linear coupling strength $\lambda$ and the phonon frequency 
$\Omega$. Figs. \ref{Fig:5}a and \ref{Fig:5}b show $\chi_C(\pi,\pi)$ and 
$\chi_{SC}$, respectively, for $\Omega = t$, $3t/2$, $2t$, and $3t$ and 
$\lambda = 0.25$. Figs. \ref{Fig:5}c and \ref{Fig:5}d show similar results 
for $\lambda = 0.5$. The results follow the trends we have discussed. 
For $\xi = 0$ the CDW correlations increase with increasing linear coupling 
or with decreasing phonon frequency, consistent with 
prior work \cite{ScalettarPRB1989}. In all cases, however, the CDW correlations 
are suppressed for increasing non-linear interaction strengths. 
The weakening of the effective linear coupling and suppression of the 
CDW ordered phase by the non-linear interaction is therefore a generic result.    

In summary, we have examined a non-linear Holstein model on a two-dimensional
square lattice and at finite temperatures and carrier concentrations using
determinant quantum Monte Carlo.  
The competition between CDW and superconducting correlations was re-examined
as a function of the non-linear $e$-ph interaction strength.
The primary effect of the non-linear coupling is a dramatic suppression of the
CDW correlations that dominate the linear model. A less pronounced
effect was observed for the superconducting correlations. 
These effects are attributed to a combined hardening of the phonon frequency 
and renormalization of the effective linear coupling by the non-linear terms. 
The net result is an overall reduction in the $e$-ph interaction strength. 
We therefore conclude that the non-linear interactions in the Holstein model 
will also reduce the transition temperature for phonon-mediated superconductivity. 
Our results explicitly confirm the importance
of the non-linear interactions at finite carrier concentrations, as proposed in
Ref. \onlinecite{AdolphsEPL2013} and show that the undressing of the polaron by the non-linear 
interactions is generic. 

Our results have implications for pump-probe experiments aimed at studying the
strength of the $e$-ph interaction. For example, the higher order interaction
terms can become important in such experiments if the external field drives the
lattice to large displacements, even if the electron-lattice coupling at
equilibrium is weak. Therefore, if the lattice is pumped too strongly there is a
danger that the non-linearity will enter and renormalize the effective coupling
to smaller values. In doing so, one could drive the system hard enough
that they extinguish the very interactions they are trying to probe. Obviously this
will be less of an issue if the pump pulses are weak and the lattice is only 
slightly perturbed; however, these effects may become extremely important 
if the lattice is strongly pumped, as our results show that even a small 
non-linear contribution can have an order-of-magnitude impact.  
Moving forward it will be important to study the role of non-linear
electron-lattice coupling and anharmonic lattice potentials as the community
continues to study systems driven far from equilibrium.

Finally, these results call for a re-evaluation of claims of
high-T$_c$ superconductivity mediated by non-linear $e$-ph coupling \cite{HighTc}. 
We stress, however, that physics arising from the non-linear coupling is different from
anharmonic effects due to the lattice potential, which are thought to
play a key role in MgB$_2$ \cite{MgB2} and KOs$_2$O$_6$ \cite{Anharmonic}.

We thank M. Berciu and D. J. Scalapino for useful
discussions. This work is partially based upon computational resources
supported by the University of Tennessee and Oak Ridge National Laboratory’s
Joint Institute for Computational Sciences (http://www.jics.utk.edu).


\begin{thebibliography}{99}
 \bibitem{Kink}
  S. Engelsberg and J. R. Schrieffer, Phys. Rev. {\bf 131}, 993 (1963).  
 \bibitem{Polaron} 
  J. T. Devreese, Encyclopedia of Applied Physics {\bf 14}, 383 (1996).
 \bibitem{MA} 
  G. L. Goodvin, M. Berciu, and G. A. Sawatzky, Phys. Rev. B {\bf 74}, 245104 (2006). 
 \bibitem{AdolphsEPL2013}
  C. P. J. Adolphs and M. Berciu, EPL {\bf 102}, 47003 (2013). 
 \bibitem{AdolphsPRB2014}
  C. P. J. Adolphs and M. Berciu, Phys. Rev. B {\bf 90}, 085149 (2014).  
 \bibitem{Mat1}
  A. Lanzara {\it et. al.}, Nature {\bf 412}, 510 (2001).
 \bibitem{Mat2}
  W. S. Lee {\it et. al.}, Phys. Rev. Lett. {\bf 110}, 265502 (2013).
 \bibitem{Mat3}
  N. Mannella {\it et al.}, Nature, {\bf 438}, 474 (2005). 
 \bibitem{Mat4}
  M. Medarde {\it et al.}, Phys. Rev. Lett. {\bf 80}, 2397 (1998). 
 \bibitem{MgB2}
  T. Yildirim {\it et al.}, Phys. Rev. Lett. {\bf 87}, 037001 (2001).
 \bibitem{Mat5}
  S. Ciuchi and S. Fratini, Phys. Rev. Lett.{\bf 106}, 166403 (2011).
 \bibitem{SC}
  {B. Bardeen, J. Cooper, and J. R. Schrieffer}, {Phys. Rev.} {\bf 108}, 1175 (1957).
 \bibitem{CDW}
  G. Gr{\"u}ner, Rev. Mod. Phys. {\bf 60}, 1129 (1988). 
 \bibitem{Time1}
  L. Perfetti {\it et al.}, Phys. Rev. Lett. {\bf 97}, 067402 (2006). 
 \bibitem{Time2} 
  F. Schmitt {\it et al.}, Science {\bf 321}, 1649 (2008). 
 \bibitem{Time3}
  A. D. Caviglia {\it et al.}, Phys. Rev. Lett. {\bf 108}, 136801 (2012). 
 \bibitem{Time4}
  E. Papalazarou {\it et al.}, Phys. Rev. Lett. {\bf 108}, 256808 (2012). 
 \bibitem{Time5}
  L. X. Lang {\it et al.}, Phys. Rev. Lett. {\bf 112}, 207001 (2014).   
 \bibitem{Time6}
  W. Zhang {\it et al.}, Nat. Comm. {\bf 5}, 4959 (2014). 
 \bibitem{WhitePRB1989}
  S. R. White {\it et al.}, Phys. Rev. B {\bf 40}, 506 (1989).
 \bibitem{BSD}
  R. Blankenbecler, D. J. Scalapino, and R. L. Sugar, Phys. Rev. D, {\bf 24}, 2278 (1981).
 \bibitem{ScalettarPRB1989}
  R. T. Scalettar, N. E. Bickers, and D. J. Scalapino, Phys. Rev. B, {\bf 40}, 197 (1989).
 \bibitem{CreffieldEPJ2005}
  C. E. Creffield, G. Sangiovanni, and M. Capone, {Eur. Phys. J B} {\bf 44}, 175 (2005).
 \bibitem{JohnstonPRB2013}
  S. Johnston {\it et al.}, Phys. Rev. B {\bf 87}, 235133 (2013).
 \bibitem{MarsiglioPRB1990}
  F. Marsiglio, Phys. Rev. B {\bf 42}, 2416 (1990).
 \bibitem{TrivediPRL1995}
  N. Trivedi and M. Randeria, Phys. Rev. Lett. {\bf 75}, 312 (1995).
 \bibitem{HighTc}
  D. M. Newns and C. C. Tsuei, Nat. Phys. {\bf 3}, 184 (2007),
 \bibitem{Anharmonic}
  J. Chang, I. Eremin, and P. Thalmeier, New J. Phys. {\bf 11}, 055068 (2009). 
\end{thebibliography}
\end{document}